\documentclass[10pt, conference, compsocconf]{IEEEtran}
\usepackage{mathptmx}
\usepackage[dvips]{graphicx}
\graphicspath{{figures/}}
\DeclareGraphicsExtensions{.eps}
\usepackage{color}
\definecolor{grey}{rgb}{0.95, 0.95, 0.95}
\usepackage[final]{listings}
\usepackage{array}
\usepackage{textcomp}
\usepackage[T1]{fontenc}
\usepackage{tikz}
\usetikzlibrary{calc,arrows,patterns}
\tikzset{>=latex}
\usepackage{verbatim}
\usepackage{flushend}

\begin{document}

\title{LIKWID Monitoring Stack: A flexible framework enabling job specific performance monitoring for the masses}

\author{\IEEEauthorblockN{Thomas R\"ohl, Jan Eitzinger, Georg Hager, Gerhard Wellein}
\IEEEauthorblockA{Erlangen Regional Computing Center (RRZE)\\
University of Erlangen-Nuremberg\\
Erlangen, Germany\\
Email: thomas.roehl|jan.eitzinger|georg.hager|gerhard.wellein@fau.de}}
\maketitle
\begin{abstract}
    System monitoring is an established tool to measure the utilization and
    health of HPC systems. Usually system monitoring infrastructures make no
    connection to job information and do not utilize hardware performance
    monitoring (HPM) data. To increase the efficient use of HPC systems
    automatic and continuous performance monitoring of jobs is an essential
    component. It can help to identify pathological cases, provides instant
    performance feedback to the users, offers initial data to judge on the
    optimization potential of applications and helps to build a statistical
    foundation about application specific system usage. The LIKWID monitoring
    stack is a modular framework build on top of the LIKWID tools library. It aims
    on enabling job specific performance monitoring using HPM data, system metrics and
    application-level data for small to medium sized commodity clusters. Moreover, it
    is designed to integrate in existing monitoring infrastructures to speed up the
    change from pure system monitoring to job-aware monitoring.
\end{abstract}

\section{Introduction}\label{sec:intro}

Large HPC systems are expensive, and so is their operation,
which makes their efficient use a crucial goal.
However, those systems are complex with regard to hardware architectures, network topologies, tool chains and software environments.
Particularly in academic computing centers there is a vast variety of applications with very different hardware demands.
Furthermore, small- to medium-sized HPC sites tend to have very limited resources for user support and application performance tuning. For them, it is not feasible to manually ensure an efficient use of the systems.
\emph{Job-specific performance monitoring} can  ease this burden.
Desirable features of such a monitoring infrastructure are:
\begin{itemize}
\item Detect pathological job behavior (idle, exceeded memory capacity, unreasonable strong scaling).
\item Give the users instant feedback and live performance data of their jobs.
\item Idendify applications with significant optimization potential and provide initial profiling data.
\item Enable application-specific statistical performance analysis of system usage for optimizing operational settings and guiding future procurements.
\end{itemize}
Despite its obvious advantages there is negligible proliferation of job-specific performance monitoring tools in small- to medium-sized academic HPC centers.
A possible reason for this may be the intricacy of the software and hardware environments on HPC systems.
Monitoring systems require a node level data acquisition agent, an infrastructure for collecting and storing the data,
some means of connecting job information to monitoring data and finally a front-end for analyzing, processing and presenting the data in a meaningful and accessible way.
For every of those components many solutions are already available and many systems already have some kind of monitoring infrastructure in place.
To come up with a generic tool integrating with or extending existing solutions is difficult.
Also the task to choose and interpret a reasonable set of performance counter events for a variety of processor and accelerator architectures is a hard problem.
On the system administrator side there are concerns that continuous and system-wide performance monitoring and the collection of the produced data might cause significant overhead.

The LIKWID Monitoring Stack tries to address these concerns and issues.
First, it tries to keep the effort for integrating the monitoring solution into existing infrastructures as low as possible.
Second, the communication protocol inside the whole system (HTTP) is commonly available on all machines.
Moreover, the stack is independent of the job scheduler software.
The integration of hardware performance measurements (HPM), derived HPM metrics, system- and application-level data provides sufficient information for further evaluation.
The analysis can be performed online to detect badly behaving jobs directly for instant user feedback or offline for in-depth analysis.
The web front-end for visualizing the job data is automatically updated .
Its views are templated and allow a variety of visualization options like graphs, histograms, pie charts and more.
Special views for administrators give an overview of running jobs.

This paper is organized as follows: In Section \ref{sec:related} related work
and the compatibility of LMS with existing solutions is discussed. The LMS
architecture is introduced in Section \ref{sec:infrastructure}, while Section
\ref{sec:appmon} describes the application-level monitoring feature. Finally, Section
\ref{sec:analysis} explains the analysis methods performed on the job data sources.

\section{Related Work}\label{sec:related}

Job-specific performance monitoring using hardware performance counting facilities is in the focus of tool developers since the early 2000s.
Especially large HPC centers, such as the National Labs in the US and the Gauss Supercomputing Centers in Germany, are active in this field.
Examples are the \emph{NWPerf} Monitoring tool developed at PNNL \cite{mooney2004nwperf}, efforts at LANL \cite{moore2015monitoring}, the \emph{Periscope} tool \cite{benedict2009automatic} and its successor \emph{PerSyst} \cite{guillen2014persyst} developed at LRZ Garching, and the \emph{HOPSA} tool collection \cite{mohr2013hopsa} developed at J\"ulich Supercomputing Centre.
Most of this work focuses on the technical challenges in scaling out a measurement infrastructure on large machines without disturbing production runs while keeping the generated data volume under control.
The only commercial vendor offering a built-in HPM job monitoring with user feedback is Cray.
Many of these solutions are site- or vendor-specific and are thus not easy to deploy at other sites.
A solution that also targets small- to medium-sized clusters is \emph{TACC Stats} \cite{evans2014comprehensive}, which is also used as part of the larger \emph{XDMoD} project \cite{Palmer2015}.
Recent and current efforts include the \emph{FEPA} project \cite{carias2015knowledge}, from which also the approach presented in this paper originates, and the just-started \emph{ProfiT-HPC} \cite{profit-hpc}.

LMS targets small- to medium-sized commodity clusters.
It employs a simple architecture for collecting the data using a time-series database.
The portability with regard to HPM events is abstracted by using the performance groups offered by the LIKWID library \cite{Treibig:2011,likwid-web}.
Due to simple standardized interfaces, all its components can be used also as standalone tools.
This eases the integration of LMS in existing monitoring infrastructures.

\section{Monitoring Architecture}\label{sec:infrastructure}

\begin{figure}[tbp]
\begin{center}
\includegraphics[width=.5\textwidth]{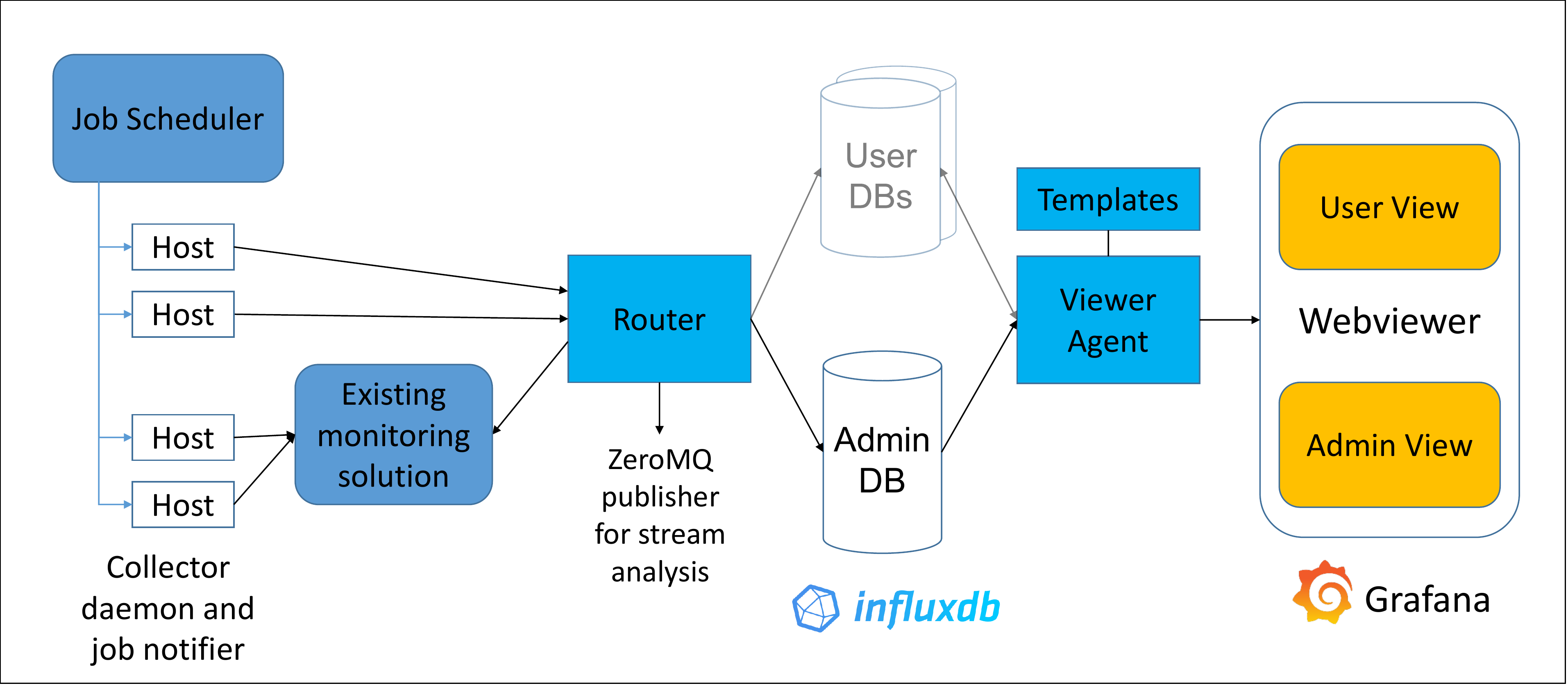}
\caption{Architecture of the LIKWID Monitoring Stack (LMS). The components are loosely coupled and can be used separately. All communication from the hosts up to the databases is handled via the InfluxDB line protocol. The Router tags the metrics (e.g., with job identifiers) and allows the duplication in user-specific databases. The Viewer Agent creates dashboards for Grafana out of templates and the metrics stored in the global or user databases.}
\label{architecture}
\end{center}
\end{figure}

This section describes the monitoring infrastructure of the LIKWID Monitoring Stack (LMS) as shown in Fig.~\ref{architecture}.
Existing Open Source components will only be covered briefly.

\subsection{Host and Forward Agent}
Most metrics are  gathered from the compute nodes, but also other systems need to be monitored to capture the whole system state, such as file servers and management nodes.
For the collection of metrics and events a variety of solutions exist.
Most of them can be integrated into LMS as the only requirement is the delivery over HTTP in the InfluxDB line protocol.
The protocol was chosen because it separates the metric value(s) from the metric tags, multiple lines can be concatenated for batched transmission and it is human-readable for better debugging.
For our tests we used the Python-based data collection daemon \emph{Diamond} \cite{diamond-web}, cronjobs sending metrics with {\tt curl} and cronjobs supplying the metrics to \emph{Ganglia} \cite{massie2004ganglia}, where the metrics are later pulled from.
The only mandatory tag for all metrics and events is the host name which is used as key in the tag store's hash table.

In order to separate the job measurements, the compute nodes or a central management server must send signals at (de)allocation of a job to the router.
The signals are piggybacked with tags, which are attached to all measurements and events from the participating hosts during the job's runtime.


\subsection{Metrics Router}

The metrics router is responsible for tagging the data with job identifiers and additional information, and for forwarding it to the database.
The router mimics the HTTP interface of an InfluxDB database plus an endpoint for job start and end signals.
For data that needs to be pulled from other sources, like the XML-interface of Ganglia's monitoring daemon {\tt gmond}, a pulling proxy can push the data into the router.
Received signals are forwarded into the database to be used later as annotations in the graphs.
All metrics are enriched with the tags from the tag store (if any) before they are forwarded to the database system.
Since all received metrics contain the hostname tag, the hostname can be used as key for the hash table of the tag store.
If configured, the router duplicates the metrics and store them in another storage location, e.g., a per-user database.
In order to attach other tools like aggregators and stream analyzers to the router, the meta information (job starts, tags, ...) and the metrics can be published via ZeroMQ \cite{zeromq-web}.

\subsection{Database Back-end}

For our setup we have chosen the InfluxDB time-series database.
It can handle floating-point data as well as strings as input values representing metrics and events.
Other database solutions can be attached, but depending on their feature set, metrics and events may have to be stored separately and handling time-series data be done explicitly due to the lack of a time format.

\subsection{Dashboard Agent}

The chosen web framework for visualization in LMS is \emph{Grafana} \cite{grafana-web}.
Although its main purpose is the presentation of real-time data, its rich feature set can be exploited to set up all the required data views.
Grafana is not configured manually but we developed a Grafana Agent that generates the dashboards out of templates, based on available databases and the metrics in them.
Most system metrics are the same for all compute nodes, but with application-level monitoring (see Sect.~\ref{sec:appmon}) additional metrics may be available.

\begin{figure}[tbp]
\begin{center}
\includegraphics[width=.5\textwidth]{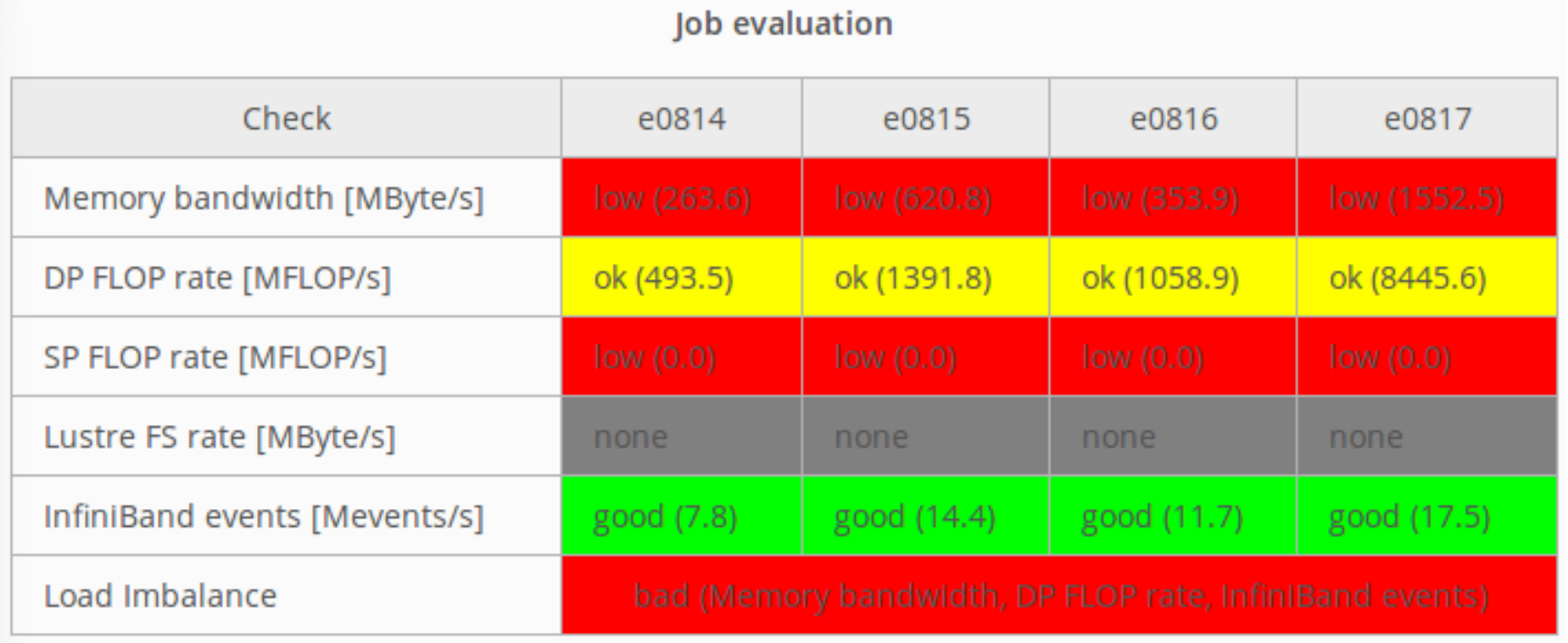}
\caption{Output of the online job evaluation with data from the start of the job until the loading of the Grafana dashboard. The four rightmost columns represent the nodes on which the job is running.}
\label{job-evaluation}
\end{center}
\end{figure}

Based on the hostnames participating in the job, the agent selects the templates for dashboard creation.
The dashboard templates can be created in Grafana, and the resulting JSON-based configuration is saved in the template location.
The dashboard, row and panel templates are combined to a full dashboard and some settings are adjusted for the current job.
As a header, analysis results of the job are presented to see badly behaving jobs on the initial view, as shown in Fig.~\ref{job-evaluation}.
The main view for administrators contains all currently running jobs with small thumbnails of the job's graphs and further information.



\section{Application-level monitoring}\label{sec:appmon}

Being able to digest application-level data adds considerable versatility to a monitoring framework.
Existing annotation libraries such as \emph{Caliper} \cite{boehme2016caliper} have a rich feature set,
but for most purposes it is sufficient to provide \emph{values} and \emph{events}.
We created the lightweight {\tt libusermetric} library, which buffers and sends batched messages using the InfluxDB line protocol.
Default tags can be specified and added to each message.
Besides metric name, value, default tags and time stamp, arbitrary tags can be supplied, such as a thread identifier.

In order to cover common use cases we created a number of automatically preloadable libraries that provide monitoring data in an application-transparent way.
The libraries overload common functions for thread affinity and data allocation.
Moreover, further information is planned to be gathered through the tooling interfaces of common parallelization solutions like MPI or OpenMP.
For use in batch scripts, a command line application can send metrics and events from the shell.

\begin{figure}[tbp]
\begin{center}
\includegraphics[width=.23\textwidth]{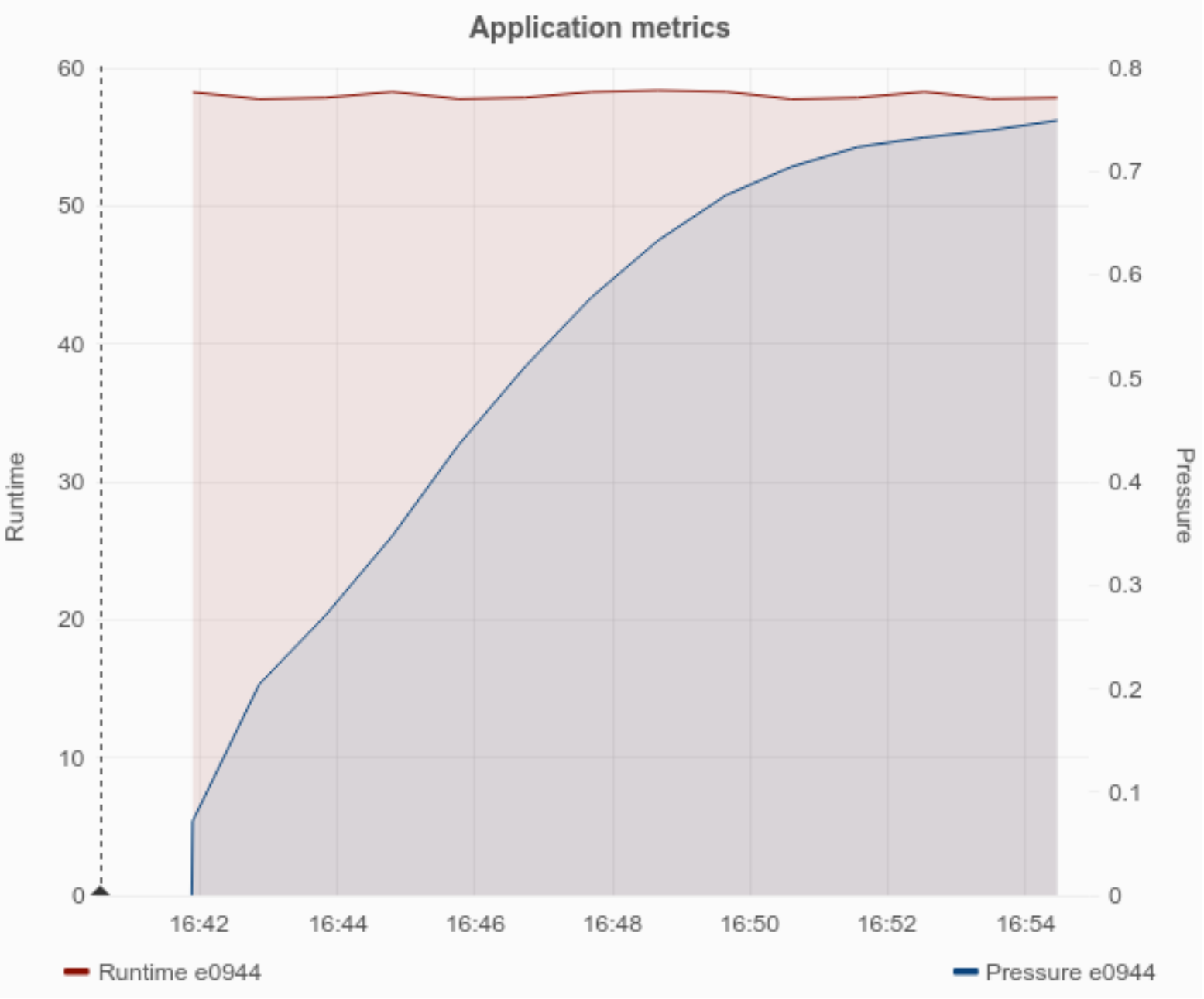} \includegraphics[width=.23\textwidth]{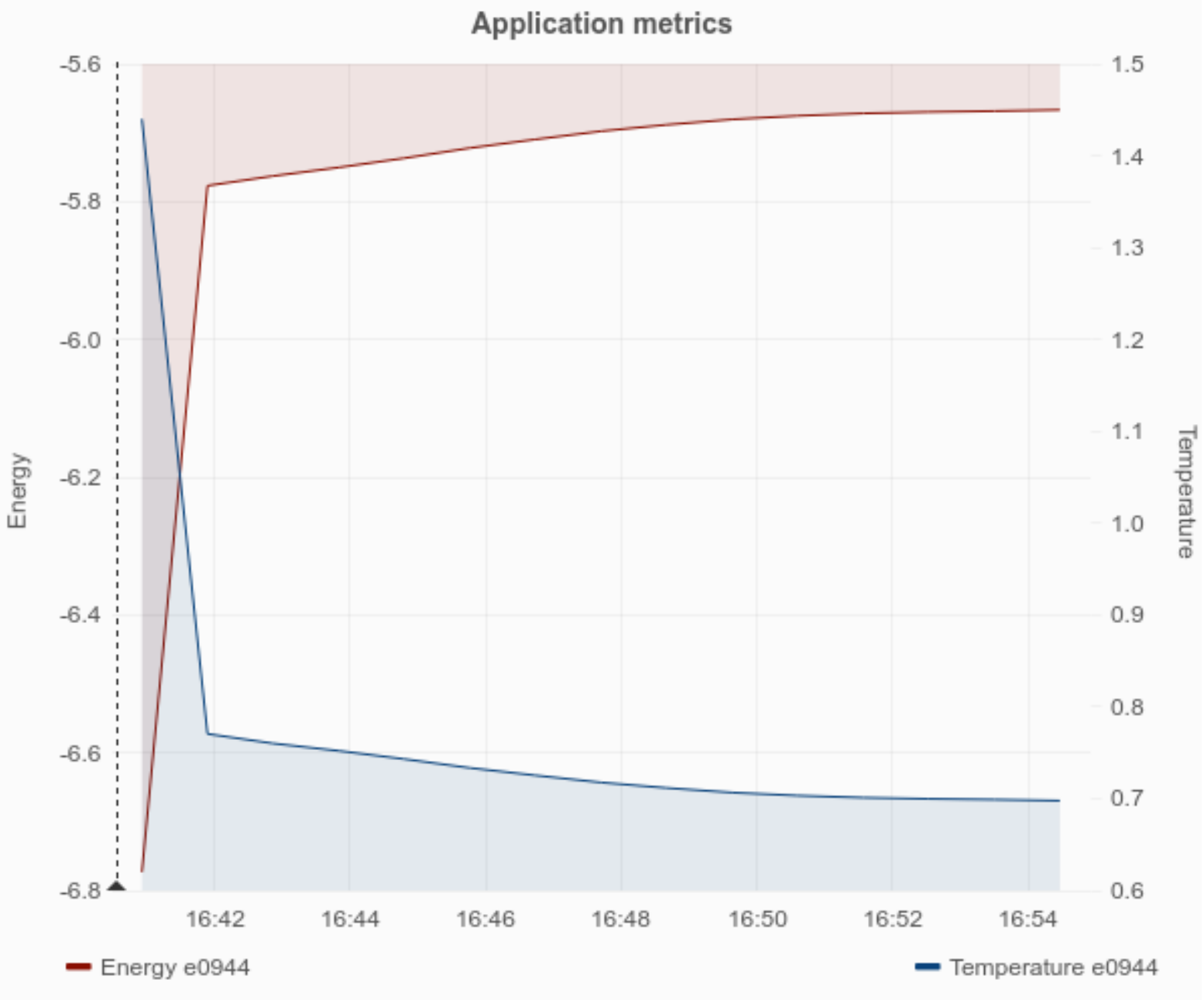}
\caption{With code annotations the user can monitor the progress of their job. This enables them to react if the application behaves unexpectedly. Left: Runtime of $100$ iterations and the pressure of the molecules in Mantevo's miniMD proxy application. Right: Energy and temperature. The events at the beginning and end of the application run are sent with the {\tt libusermetric} command line tool.}
\label{application-monitoring}
\end{center}
\end{figure}

A typical use case for application data monitoring is shown in Fig.~\ref{application-monitoring}: Four metrics (runtime for 100 iterations, pressure, temperature and energy) of a run with Mantevo's miniMD proxy application are displayed versus the runtime.
Moreover, two events are supplied before starting and after finishing the execution of miniMD and are represented as dark dashed lines.
%


\section{Data analysis methodology}\label{sec:analysis}

A crucial task for any monitoring system is the definition of relevant metrics and their systematic interpretation. Metrics are usually based on raw event counts from which derived values, such as rates and bandwidths, can be calculated.
The metrics used in LMS are drawn from a combination of system-level, application-level and hardware performance counter data sources.
The initial analysis focuses on elementary resource utilization using the following data:
\begin{itemize}
    \item CPU load
    \item IPC and floating-point rates
    \item allocated memory size
    \item memory bandwidth
    \item network I/O
    \item file I/O
\end{itemize}
\begin{figure}[tbp]
\begin{center}
\includegraphics[width=.5\textwidth]{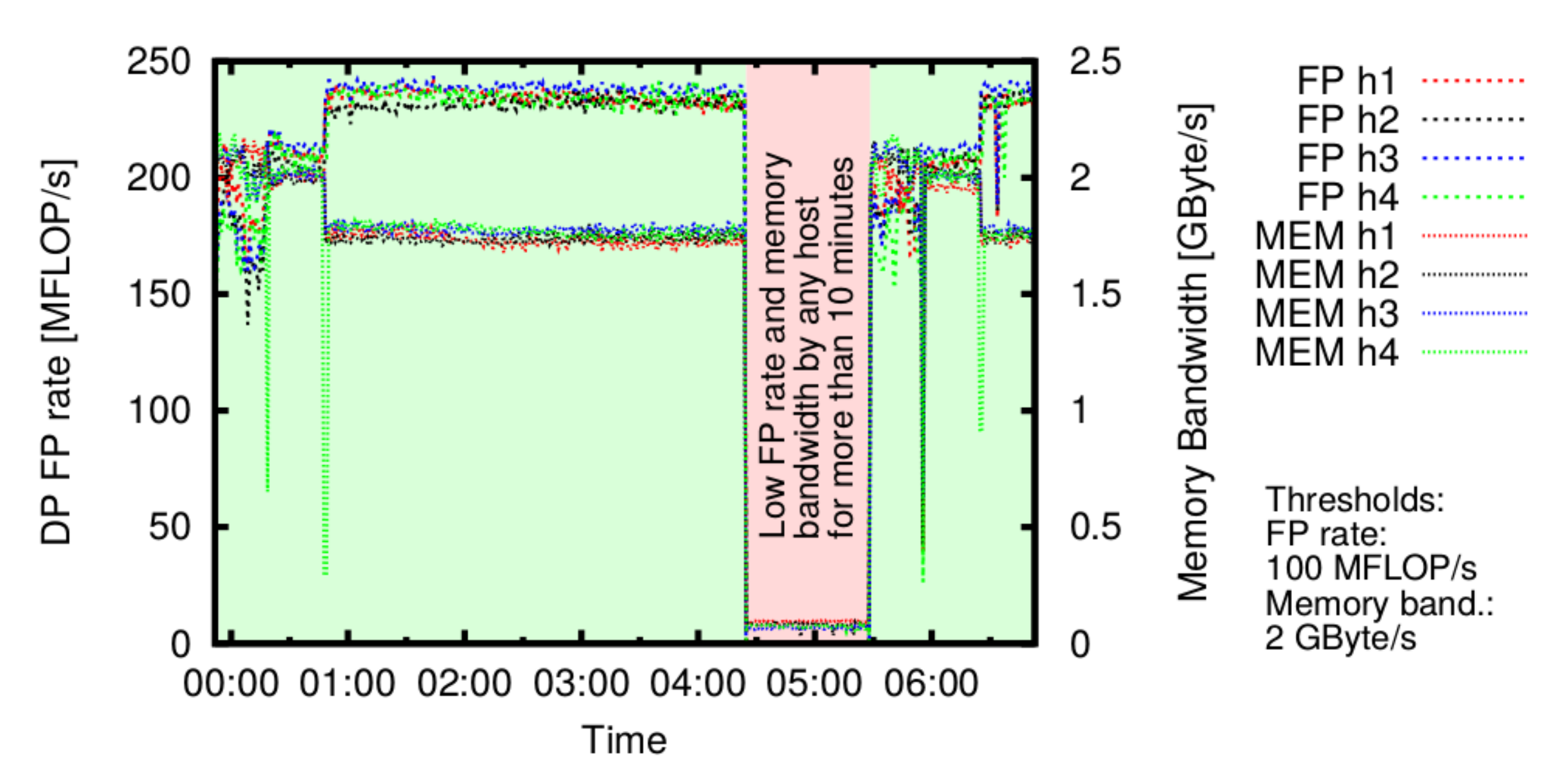}
\caption{Timeline of the DP FP rate and memory bandwidth of an four-node (h1, h2, h3 and h4) job run revealing a longer break in computation with FP rate and memory bandwidth below thresholds for more than 10 minutes.}
\label{job-with-break}
\end{center}
\end{figure}

The metrics derived from hardware performance counters are based on the event sets provided by the LIKWID performance groups.
The detection of pathological jobs is based on simple rules for the resource utilization metrics using thresholds and timeouts like in Fig.~\ref{job-with-break}.
For marking applications with significant optimization potential we use the performance pattern systematic initially described in \cite{Treibig:2013} and later refined as part of the FEPA project using a decision tree \cite{carias2015knowledge}.




\section{Conclusion}\label{sec:conclusion}

In this paper we presented the LIKWID Monitoring Stack (LMS).  LMS is
an effort to create a flexible set of Open Source components connected
by simple interface scripts.  The components can be used as a complete
stack, standalone or in parts if integrated in an existing
infrastructure.  LMS is designed for deployment in small- to
medium-sized commodity cluster systems where an intricate data collection
infrastructure is not required.  It offers live job performance
profiling on the system level or per user.  The system allows
to detect pathological jobs based on resource utilization and uses a
performance pattern decision tree for judging the optimization
potential of applications.  The LMS is currently in beta status,
but it is being tested at several HPC computing centers in Germany.
\bibliography{IEEEabrv,hpcmspa2017}
\bibliographystyle{IEEEtran}
\end{document}